\documentclass{article}
\newcommand{\ket}[1]{\left | \, #1 \right \rangle}
\newcommand{\bra}[1]{\left \langle #1 \, \right |}
\newcommand{\proj}[1]{\ket{#1}\bra{#1}}

\begin{document}

\bigskip
\centerline{\Large \bf Against Quantum Noise}
\bigskip\bigskip
\centerline{\large Artur Ekert
\footnote{\tt ekert@physics.ox.ac.uk} 
and Chiara Macchiavello 
\footnote{\tt chiara@mildred.physics.ox.ac.uk.
Also at: Dipartimento di Fisica Generale ``A. Volta'', Via Bassi 6, 
27100 Pavia, Italy}}
\medskip
\centerline{\it Clarendon Laboratory, University of Oxford, Oxford OX1
3PU, U.K.}
\bigskip

\begin{abstract}
This is a brief description of how to protect quantum states from dissipation and
decoherence that arise due to uncontrolled interactions with the environment. We
discuss recoherence and stabilisation of quantum states based on two techniques known
as ``symmetrisation" and ``quantum error correction". We illustrate our
considerations with the most popular quantum-optical model of the system-environment
interaction, commonly used to describe spontaneous emission, and  show the benefits
of quantum error correction in this case.
\end{abstract}
\parskip=5pt 
\section{Introduction}
\label{s:int}

Suppose we want to transmit or store a block of $l$ qubits (i.e. two-state
quantum systems) in a noisy environment.  Here `noisy' means that each qubit
may become entangled with the environment. Thus due to spurious interactions
with the environment the actual state of the $l$ qubits, described by a
density operator $\rho(t)$, will differ from the original state $\ket{\Psi}$.
This deviation can be quantified by the fidelity
\begin{equation}
F(t)=\bra{\Psi}\rho(t)\ket{\Psi} =1-\epsilon(t).
\end{equation}
In order to maximize this fidelity we may try all sorts of tricks ranging
from the most obvious one i.e. isolating the qubits from the environment to
more sophisticated methods such as ``symmetrisation" \cite{Deutsch1993,Barenco
etal}, ``purification" \cite{Bennett etal,QPA}, and ``quantum error
correction" \cite{Shor}. 
The last method seems to be the most popular one at
the moment and relies on encoding the state of $l$ qubits into a set of $n$
qubits and trying to disentangle a certain number of qubits from the
environment after some period of time. In the following we describe, very
briefly, how some of these techniques work.

We will assume that in the block of $l$ qubits each qubit is coupled to a different environment.
This is a perfectly reasonable assumption, which is valid if the coherence length of the
environment/reservoir is less than the spatial separation between the qubits~\cite{palma}, and
introduces a great deal of simplifications to the calculations. Basically it allows us
to view any dissipation of $l$ qubits as a set of independent dissipations of $l$
single qubits (i.e. we ignore collective phenomena such as superradiance etc.).

The qubit--environment interaction leads to the qubit--environment
entanglement, which in its most general form is given by
\begin{eqnarray}
\ket{0}\ket{R} & \longrightarrow & \ket{0}\ket{R_{00}(t)} +
\ket{1}\ket{R_{01}(t)},\\ \ket{1}\ket{R} & \longrightarrow &
\ket{0}\ket{R_{10}(t)} + \ket{1}\ket{R_{11}(t)},
\label{dissipation}
\end{eqnarray}
where states of the environment $\ket{R}$ and $\ket{R_{ij}}$ are
neither normalised nor orthogonal to each other (thus we have to take
additional care at the end of our calculations and normalise the final
states). The r.h.s. of the formulae above can also be written in a
matrix form as
\begin{equation}
\left( \begin{array}{cc}
\ket{R_{00}} & \ket{R_{01}} \\
\ket{R_{10}} & \ket{R_{11}} 
\end{array}
\right)
\left( \begin{array}{c}
\ket{0}\\ \ket{1}\end{array}\right ),
\end{equation}
and the 2 $\times$ 2 matrix can can be subsequently decomposed into
some basis matrices e.g. into the unity and the Pauli matrices
\begin{equation}
\ket{R_0}1+\ket{R_1}\sigma_x + i\ket{R_2}\sigma_y + \ket{R_3}\sigma_z,
\end{equation}
where $\ket{R_0}= (\ket{R_{00}}+\ket{R_{11}})/2$, $\ket{R_3}=
(\ket{R_{00}}-\ket{R_{11}})/2$, $\ket{R_1} = (\ket{R_{01}}+\ket{R_{10}})/2$,
and $\ket{R_2} =(\ket{R_{01}}-\ket{R_{10}})/2$.  Thus the qubit initially
in state $\ket{\Psi}$ will evolve as
\begin{equation}
\ket{\Psi}\ket{R} \longrightarrow \sum_{i=0}^3\sigma_i\ket{\Psi}\ket{R_i}
\label{action}
\end{equation}
becoming entangled with the environment (we have relabelled the unity
operator and the Pauli matrices $\{1, \sigma_x, \sigma_y, \sigma_z\}$
respectively as $\{\sigma_0, \sigma_1, \sigma_2, \sigma_3\}$). Its
fidelity with respect to the initial state $\ket{\Psi}$ evolves as
\begin{equation}
F(t) = \sum_{i,j} \bra{\Psi}\sigma_i\ket{\Psi}
\bra{\Psi}\sigma_j\ket{\Psi}\bra{R_j(t)}R_i(t)\rangle .
\label{fidelity}
\end{equation}
 
The formula (\ref{action}) describes how the environment affects any
quantum state of a qubit and shows that a general qubit--environment interaction
can be expressed as a superposition of unity and Pauli operators acting
on the qubit. As we will see in the following, in the language of error 
correcting codes this means that the qubit state is evolved into a superposition
of an error-free component and three erroneous components, with errors
of the $\sigma_x$, $\sigma_y$ and $\sigma_z$ type.

We can carry on this description even if the qubit itself is not in a pure 
state $\ket{\Psi}$ but is
entangled with some other qubits. For example, if in a three qubit
register initially in state $\ket{\tilde\Psi} =
\ket{0}\ket{0}\ket{0}-\ket{1}\ket{1}\ket{1}$ the second qubit
interacted with its environment then the state of the register at some
time $t$ is given by
\begin{equation}
\sum_{i=0}^3\sigma_i^{(2)} \ket{\tilde\Psi}\ket{R_i (t)} =
\sum_{i=0}^3(\ket{0}(\sigma_i\ket{0})\ket{0} -
(\ket{1}(\sigma_i\ket{1})\ket{1})\ket{R_i (t)},
\label{int}
\end{equation}
where the superscript $(2)$ reminds us that the Pauli operators act
only on the second qubit. We can then say that the second qubit was affected
by quantum errors which are represented by the Pauli operators
$\sigma_i$. Errors affecting classical bits can only change their
binary values ($0\leftrightarrow 1$), in contrast
quantum errors operators $\sigma_i$ acting on qubits can change their
binary values ($\sigma_x$), their phases ($\sigma_z$) or both
($\sigma_y$).

In general, a batch of $n$ qubits initially in some state
$\ket{\tilde\Psi}$, each of them interacting with different 
environments, will evolve as
\begin{equation}
\prod_{k=1}^n\sum_{i=0}^3\sigma_i^{(k)} \ket{\tilde\Psi}\ket{R_i^{(k)} (t)}\;,
\end{equation}
namely multiple errors of the form 
$\sigma_i\otimes\sigma_j\cdot\cdot\cdot\otimes\sigma_k$ may occur, affecting
several qubits at the same time.

So much about unwelcome dissipation, what about remedies?

\section{Stabilization via symmetrisation}

The first proposed remedy was based on a symmetrisation procedure
\cite{Deutsch1993}. The basic idea is as follows. Suppose you have a quantum
system, you prepare it in some initial state $\ket{\Psi_i}$ and you
want to implement a prescribed unitary evolution $\ket{\Psi (t)}$ or
simply you want to preserve $\ket{\Psi_i}$ for some period of time
$t$. Now, suppose that instead of a single system you can prepare $R$
copies of $\ket{\Psi_i}$ and subsequently you can project the state of
the combined system on the symmetric subspace i.e. the subspace
containing all states which are invariant under any permutation of the
sub-systems. The claim is that frequent projections on the symmetric
subspace will reduce errors induced by the environment. The intuition
behind this concept is based on the observation that a prescribed
error-free storage or evolution of the $R$ independent copies starts
in the symmetric sub-space and should remain in that
sub-space. Therefore, since the error-free component of any state
always lies in the symmetric subspace, upon successful projection it
will be unchanged and part of the error will have been removed. Note
however that the projected state is generally not error--free since
the symmetric subspace contains states which are not of the simple
product form $\ket{\psi}\ket{\psi}\ldots\ket{\psi}$. Nevertheless it
has been shown that the error probability will be suppressed by a
factor of $1/R$ \cite{Barenco etal}.

We illustrate here this effect in the simplest case of two qubits.
The projection into the symmetric subspace is performed in this case
by introducing the symmetrisation operator:
\begin{equation}
\label{symop2}
S= \frac{1}{2} (P_{12}+P_{21})\;,
\end{equation}
where $P_{12}$ represents the identity and $P_{21}$ the permutation
operator which exchanges the states of the two qubits. The
symmetric--projection of a pure state $\ket{\Psi} $ of two qubits is
just $S\ket{\Psi}$, which is then renormalised to unity. It follows
that the induced map on mixed states of two qubits (including
renormalisation) is:
\begin{equation}
\label{symproj2}
\rho_1 \otimes \rho_2 \longrightarrow
  \frac{S(\rho_1 \otimes\rho_2 ) 
S^{\dagger}}{\mbox{Tr}{S(\rho_1 \otimes\rho_2 )
S^{\dagger}}}
\end{equation}
The state of either qubit separately is then obtained by partial trace
over the other qubit.

Consider for example the symmetric projection of $\rho\otimes \rho$
followed by renormalisation and partial trace (over either qubit) to
obtain the final state ${\rho_s}$ of one qubit, given that the
symmetric-projection was successful. A direct calculation based on
(\ref{symproj2}) yields:
\begin{equation}
\label{twid}
\rho \mapsto {\rho_s}= \frac{\rho +\rho^2}{\mbox{Tr}{(\rho +\rho^2 ) }}
\end{equation}
For any mixed state $\xi$ of a qubit the expression $\mbox{Tr}{\xi ^2
}$ provides a measure of the purity of the state, ranging from
$\frac{1}{4}$ for the completely mixed state $I/2$ (where $I$ is the
unit operator) to 1 for any pure state. From (\ref{twid}) we get
\begin{equation}
\mbox{Tr}{{\rho_s}^2} > \mbox{Tr}{\rho ^2}
\end{equation}
so that ${\rho_s}$ is {\em purer} than $\rho$.  This illustrates that
successful projection of a mixed state into the symmetric subspace
tends to enhance the purity of the individual systems.

To be more specific, let us assume now that the two copies are
initially prepared in pure state $\rho_0 = \proj{\Psi}$ and that they
interact with independent environments. After some short period of
time $\delta t$ the state of the two copies $\rho^{(2)}$ will have
undergone an evolution
\begin{equation}
\label{dec}
\rho^{(2)}(0) = \rho_0 \otimes \rho_0 \hspace{5mm}
\longrightarrow
\hspace{5mm} \rho^{(2)}(\delta t) = \rho_1\otimes\rho_2
\end{equation}
where $\rho_i = \rho_0 + \varrho_i$ for some Hermitian traceless
$\varrho_i$.
We will retain only  terms of first order in the perturbations
$\varrho_i$ so that the overall state at time $\delta t$ is
\begin{eqnarray}
\rho^{(2)} = \rho_0 \otimes\rho_0 
+ \varrho_1 \otimes \rho_0 +\rho_0 \otimes \varrho_2  
+ O(\varrho_1 \varrho_2 )\;.
\label{decst}
\end{eqnarray}

We can calculate the average purity of the two copies before symmetrisation by calculating the
average trace of the squared states:
\begin{equation} \label{this}
\frac{1}{2} \sum_{i=1}^2 \mbox{Tr}( (\rho_0+\varrho_i)^2) = 1 + 2
 \mbox{Tr}(\rho_0\tilde\varrho ),
\end{equation}
where $\tilde\varrho=\frac{1}{2}(\varrho_1+\varrho_2)$.
Note that $\mbox{Tr}({\rho_0 \tilde\varrho})$ is negative, so that the
expression above does not exceed 1.
After symmetrisation each qubit is in state
\begin{equation} 
\rho_s=[1- \mbox{Tr}(\rho_0\tilde\varrho )] \rho_0 +\frac{1}{2}\tilde\varrho
+\frac{1}{2}(\rho_0\tilde\varrho+ \tilde\varrho\rho_0)
\end{equation}
and has purity
\begin{equation}
  \mbox{Tr}(\rho_s^2)=1 + \mbox{Tr}(\rho_0\tilde\varrho).
\end{equation}
Since $\mbox{Tr}{\rho_s^2 }$ is closer to 1 than (\ref{this}),
the resulting symmetrised
system $\rho_s$ is left in a purer state.

Let us now see how the fidelity changes by applying the symmetrisation
procedure. The average fidelity before symmetrisation is
\begin{equation} \label{fid1}
F_{bs}=\frac{1}{2} \sum_i \bra{\Psi}\rho_0 +\varrho_i\ket{\Psi} =
1+ \bra{\Psi} \tilde\varrho \ket{\Psi}\;,
\end{equation}
while after successful symmetrisation it takes the form
\begin{equation}  
F_{as}=\bra{\Psi}\rho_s \ket{\Psi} = 
1+ \frac{1}{2} \bra{\Psi}\tilde\varrho \ket{\Psi}\;.
\end{equation}
The state after symmetrisation is therefore closer to the initial state
$\rho_0$.

For the generic case of $R$ copies the purity of each qubit after 
symmetrisation is given by \cite{Barenco etal}
\begin{equation}
  \mbox{Tr}(\rho_s^2)=1 + 2\frac{1}{R} \mbox{Tr}(\rho_0\tilde\varrho)\;,
\label{purR}
\end{equation}
where now $\tilde\varrho=\frac{1}{R}\sum_{i=1}^R\varrho_i$,
and the fidelity takes the form
\begin{equation}  \label{fid2}
\bra{\Psi}\rho_s \ket{\Psi} = 1+ \frac{1}{R} \mbox{Tr}
({\rho_0 \tilde\varrho })\;.
\end{equation}

Formulae (\ref{purR}) and (\ref{fid2}) must be compared with the 
corresponding ones before symmetrisation, i.e. (\ref{this}) and (\ref{fid1}).
As we can see,  $\rho_s$ approaches the unperturbed state 
$\rho_0$ as $R$ tends to infinity.
Thus by choosing $R$ sufficiently large and the rate of symmetric projection 
sufficiently high, the residual error at the end of a computation can, 
in principle, be controlled to lie within any desired small tolerance.

The efficiency of the symmetrisation procedure depends critically on
the probability that the state of the $R$ qubits is successfully
projected into the symmetric subspace. It has been shown that if the
projections are done frequently enough, then the cumulative
probability that they all succeed can be made as close as desired to
unity.  This is a consequence of the fact that the fidelity of the
state of the $R$ computers with respect to the corresponding error
free state for small times $\delta t$ has a parabolic behaviour (see
section \ref{s:dyn}). Therefore the probability of successful projection, which
is unity at the initial time, begins to change only to second order in
time.  If we project $n$ times per unit time interval, i.e. we choose
the time interval between two subsequent projections $\delta t = 1/n$,
then the cumulative probability that all projections in one unit time
interval succeed is given by
\begin{eqnarray} 
[1-k(\delta t)^2 ]^n = (1-\frac{k}{n^2})^n \rightarrow 1
\mbox{ as } n\rightarrow \infty\;.
\end{eqnarray} 
Here $k$ is a constant depending on the rate of rotation of the state
out of the symmetric subspace.  This effect is known as the ``quantum
watch-dog effect'' or the ``quantum Zeno effect".

\section{Quantum encoding and decoding}
\label{s:enc-dec}

The idea of protecting information via encoding and decoding lies at
the foundations of the classical information theory. It is based on a
clever use of redundancy during the data storage or transmission. For
example, if the probability of error (bit flip) during a single bit
transmission via a noisy channel is $p$ and each time we want to send
bit value 0 or 1 we can encode it by a triple repetition i.e. by
sending 000 or 111. At the receiving end each triplet is decoded as
either zero or one following the majority rule - more zeros means 0,
more ones means 1. This is the simplest error correcting protocol
which allows to correct up to one error.

In the triple repetition code the signalled bit value is recovered
correctly both when there was no error during the transmission of the
three bits, which happens with probability $(1-p)^3$, and when there
was one error at any of the three locations, which happens with
probability $3p(1-p)^2$. Thus the probability of the correct
transmission (up to the second order in $p$) is $1-3p^2$ i.e. the
probability of error is now $3p^2$, which is much smaller when
compared with the probablity of error without encoding and decoding
$p$ ($p\ll 1$). 
This way we can trade the probability of error in the
signalled message for a number of transmissions via the channel. 
In our example the reduction of the error rate
from $p$ to $3p^2$ required to send three times more bits.
If sending each bit via the channel costs us money we
have to decide what we treasure more, our bank account or our
infallibility. The triple repetition code encodes one bit into three
bits and protects against one error, in general we can construct codes
that encode $l$ bits into $n$ bits and protect against $t$ errors. The
best codes, of course, are those which for a fixed value $l$ minimize
$n$ and maximize $t$.

Quantum error correction which protects quantum states is a little bit
more sophisticated simply because the bit flip is not the only
''quantum error" which may occur, as we have seen in the previous
sections. Moreover, the decoding via the majority rule does not
usually work because it may involve measurements which destroy quantum
superpositions. Still, the triple repetition code is a good starting
point to investigate quantum codes and even to construct the simplest
ones.

The simplest interesting case of the most general qubit--environment
evolution (\ref{dissipation}) is the case of decoherence \cite{Zurek1991}
where the environment effectively acts as a measuring apparatus

\begin{eqnarray}
\ket{0}\ket{R} & \longrightarrow & \ket{0}\ket{R_{00}(t)},\\ 
\ket{1}\ket{R} & \longrightarrow & \ket{1}\ket{R_{11}(t)}.
\label{decoherence}
\end{eqnarray}

Following our discussion in Section \ref{s:int} we can see that this model
leads only to dephasing errors of the $\sigma_z$ type. It turns out
that a phase flip can be handled almost in the same way as a classical
bit flip. Again, consider the following scenario: we want to store, in
a computer memory, one qubit in an {\em unknown} quantum state of the
form $\alpha\ket{0}+\beta\ket{1}$ and we know that any single qubit
which is stored in a register may, with a small probability $p$,
undergo a decoherence type entanglement with an environment
(Eq. \ref{decoherence}); in particular

\begin{equation}
(\alpha\ket{0}+\beta\ket{1})\ket{R}\longrightarrow
\alpha\ket{0}\ket{R_{00}}+\beta\ket{1}\ket{R_{11}}.
\end{equation}

Let us now show how to reduce the probability of decoherence to be of
the order $p^2$.

Before we place the qubit in the memory register we {\em encode} it:
we can add two qubits, initially both in state $\ket{0}$, to the
original qubit and then perform an encoding unitary transformation

\begin{eqnarray}
\ket{000}&\longrightarrow &\ket{C_0}
=(\ket{0}+\ket{1})(\ket{0}+\ket{1})(\ket{0}+\ket{1}),\\
\ket{100}&\longrightarrow &\ket{C_1}
=(\ket{0}-\ket{1})(\ket{0}-\ket{1})(\ket{0}-\ket{1}),
\end{eqnarray}

generating state $\alpha\ket{C_0}+\beta\ket{C_1}$. 
Now, suppose that only the second stored
qubit was affected by decoherence and became entangled with the environment:

\begin{eqnarray}
\alpha (\ket{0}+\ket{1})(\ket{0}\ket{R_{00}}
+\ket{1}\ket{R_{11}})(\ket{0}+\ket{1}) +\nonumber\\
\beta (\ket{0}-\ket{1})(\ket{0}\ket{R_{00}}
-\ket{1}\ket{R_{11}})(\ket{0}-\ket{1}),
\end{eqnarray}

which, following Eq. (\ref{int}), can be written as

\begin{equation}
(\alpha\ket{C_0} + \beta\ket{C_1})\ket{R_0} 
+ \sigma_z^{(2)}(\alpha\ket{C_0} + \beta\ket{C_1})\ket{R_3}.
\end{equation}

If vectors $\ket{C_0}$, $\ket{C_1}$, $\sigma_z^{(k)}\ket{C_0}$, and
$\sigma_z^{(k)}\ket{C_1}$ are orthogonal to each other we can try to
perform a measurement on the qubits and project their state either on
the state $\alpha\ket{C_0} + \beta\ket{C_1}$ or on the orthogonal one
$\sigma_z^{(2)}(\alpha\ket{C_0} + \beta\ket{C_1})$. The first case yields
the proper state right away, the second one requires one application
of $\sigma_z$ to compensate for the error. In this simple case one can
even find a direct unitary operation which can fix all one qubit phase
flips regardless their location. For example the transformation

\begin{eqnarray}
\ket{000}\to \ket{000} & & \ket{100} \to \ket{011}\nonumber \\
\ket{001}\to \ket{001} & & \ket{101} \to \ket{110}\nonumber \\
\ket{010}\to \ket{010} & & \ket{110} \to \ket{101}\nonumber \\
\ket{011}\to \ket{111} & & \ket{111} \to \ket{100}
\end{eqnarray}

corrects any single bit flip $0\leftrightarrow 1$ and when applied in
the conjugate basis ($\ket{0'}=\ket{0}+\ket{1}$,
$\ket{1'}=\ket{0}-\ket{1}$) it corrects any single phase flip (the bit
flips become phase flips in the new basis). The snag is that using the
scheme above we can correct up to one phase error $\sigma_z$ or we can
go to a conjugate basis and the same scheme will correct up to one
amplitude error $\sigma_x$ but it cannot correct up to one general
error, be it amplitude or phase.

To fix this problem Peter Shor in 1995 combined the phase and the
amplitude correction schemes into one constructing the following nine
qubit code \cite{Shor}:
\begin{eqnarray}
&&\ket{0}\to \frac{1}{2\sqrt{2}}(\ket{000}+\ket{111})(\ket{000}+\ket{111})
(\ket{000}+\ket{111})
\label{shor1}\\            
&&\ket{1}\to \frac{1}{2\sqrt{2}}(\ket{000}-\ket{111})(\ket{000}-\ket{111})
(\ket{000}-\ket{111})\;.
\label{shor2}
\end{eqnarray}
This code involves double encoding, first in base $\ket{0}$ and
$\ket{1}$ and then in base $\ket{0'}$ and $\ket{1'}$, and it
allows to correct up to one either bit or phase flip. It turns out
that the ability to correct both amplitude and phase errors suffices
to correct any error due to entanglement with the environment. In
other words the action of the environment on qubits can be viewed in
terms of bit and phase flips.

\section{Quantum error-correcting codes}

The original nine qubit code of Shor can be further simplified. It has
been shown that a five qubit code suffices to correct a single error
of any type. Let us now specify the conditions for the existence of
quantum error-correcting codes.

We say we can correct a single error $\sigma_i^{(k)}$ (where $i=0\ldots 3$
refers to the type of error) 
if we can find a transformation such that it maps all states
with a single error $\sigma_i^{(k)}\ket{\tilde\Psi}$ into the proper error
free state $\ket{\tilde\Psi}$:
\begin{equation}
\sigma^{(k)}_i \ket{\tilde\Psi} \longrightarrow  \ket{\tilde\Psi}
\end{equation}
To make it unitary we may need an ancilla
\begin{equation}
\sigma_i^{(k)} \ket{\tilde\Psi}\ket{0} \longrightarrow  \ket{\tilde\Psi}
\ket{a_i^k}\;.
\end{equation}

For encoded basis states of a single qubit 
$\ket{C_0}$ and $\ket{C_1}$ this implies \cite{bdws}
\begin{eqnarray}
A_k \ket{C_0 }\ket{0} & \longrightarrow & \ket{C_0}\ket{a_k}\\
A_k \ket{C_1 }\ket{0} & \longrightarrow & \ket{C_1}\ket{a_k},
\end{eqnarray}
where $A_k$ denotes all the possible types of independent errors
affecting at most one of the qubits.The above requirement leads to the
following unitarity conditions
\begin{eqnarray}
\bra{C_0} A^\dagger_k A_l\ket{C_0}&=&\bra{C_1}
A^\dagger_k A_l\ket{C_1} = \bra{a_k} a_l\rangle\;,
\label{condec}\\
\bra{C_0}A^\dagger_k A_l\ket{C_1}&=& 0 \;.
\end{eqnarray}
The above conditions are straightforwardly generalised to an arbitrary
$t$ error correcting code, which corrects any kind of transformations affecting
up to $t$ qubits in the encoded state. In this case the operators $A_k$ are
all the possible independent errors affecting up to $t$ qubits, namely 
operators of the form $\Pi_{i=1}^t\sigma_i$ acting on $t$ different qubits.
In the case of the so-called ``nondegenerate codes'' 
Eq. (\ref{condec}) takes the simple form \cite{nostro}
\begin{equation}
\bra{C_0} A^\dagger_k A_l\ket{C_0} =\bra{C_1} A^\dagger_k A_l\ket{C_1}=0\;.
\label{orth}
\end{equation}

This condition requires 
that all states which are obtained by affecting up to $t$ qubits in the 
encoded states are all orthogonal to each other, and therefore distinguishable.
This ensures that by performing suitable projections of the encoded
state we are able to detect the kind of error which occurred and ``undo'' it 
to recover the desired error free state. 
Condition (\ref{orth}), even if more restrictive than (\ref{condec}), is 
particularly useful because it allows to establish bounds on the resources 
needed in order to have efficient nondegenerate codes.
Let us assume that the initial state of $l$ qubits
is encoded in a redundant Hilbert space of $n$ qubits. 
If we want to encode $2^l$ input basis states and correct up to $t$ errors
we must choose the dimension of the encoding Hilbert space $2^n$ such that 
all the necessary orthogonal states can be accomodated.
According to Eq. (\ref{orth}), the total number of orthogonal states that we
need in order to be able to correct $i$ errors of the three types $\sigma_x$,
$\sigma_y$ and $\sigma_z$ in an $n$-qubit state is 
$3^i\left(\begin{array}{c} n \\ i \end{array}\right)$ (this is the number of
different ways in which the errors can occur).
The argument based on counting orthogonal states then leads to the following
condition
\begin{eqnarray}
2^l\sum_{i=0}^t 3^i\left(\begin{array}{c} n \\ i
\end{array}\right)\leq 2^n.\label{hamming}
\end{eqnarray}
Eq. (\ref{hamming}) is the quantum version of the Hamming bound for
classical error-correcting codes~\cite{macw}; given $l$ and $t$ it
provides a lower bound on the dimension of the encoding Hilbert space
for nondegenerate codes. Let us mention that an explicit construction for 
quantum codes for some values $(l,n,t)$ which saturate the quantum Hamming 
bound has been provided \cite{gottesman}.
It is interesting that this bound has not been beaten so far by degenerate 
codes \cite{barolo}.

The quantum version of the classical
Gilbert-Varshamov bound~\cite{macw} can be also obtained, which gives 
an upper bound on the dimension of the encoding Hilbert space for optimal
non degenerate codes:
\begin{eqnarray}
2^l\sum_{i=0}^{2t} 3^i\left(\begin{array}{c} n \\ i
\end{array}\right)\geq 2^n.\label{gvbound}
\end{eqnarray}
This expression can be proved from the observation that in the $2^n$ 
dimensional Hilbert space with a maximum number of encoded basis vectors 
(or code-vectors) ${\left | \, C^k \right \rangle}$  
any vector which is orthogonal to ${\left |
\, C^k \right \rangle}$ (for any $k$)  can be reached by applying up to
$2t$ error operations of $\sigma_x$, $\sigma_y$, and $\sigma_z$ type to
any of the $2^l$ encoded basis vectors.  Clearly all vectors which cannot be
reached in the $2t$ operations can be added to the encoded basis states 
${\left | \, C^k \right \rangle}$ as all the  vectors into which they can be
transformed by applying up to $t$ amplitude and/or phase transformations
are  orthogonal to all the others.  This situation cannot happen because
we have assumed that the number of code-vectors is maximal.  Thus the
number of orthogonal vectors that can be obtained by performing up to
$2t$ transformations on the code-vectors must be at  least equal to the
dimension of the encoding Hilbert space.

It follows from Eq. (\ref{hamming}) that a one-bit quantum error
correcting code to protect a single qubit ($l=1$, $t=1$) requires at
least $5$ encoding qubits and, according to Eq.  (\ref{gvbound}), this
can be achieved with less than $10$ qubits.  Indeed, Shor's nine qubit
code can be simplified to the seven qubit code \cite{steane}, and 
ultimately to the quantum Hamming bound \cite{bdws,lafl}. We will consider
explicitly one form of the five qubit code in Section \ref{s:beyond}.

The asymptotic form of the quantum Hamming bound (\ref{hamming}) in the
limit of large $n$ is given by 
\begin{eqnarray}
\frac{l}{n}\leq 1-\frac{t}{n}\log_2 3 -H(\frac{t}{n}),\label{hamasym}
\end{eqnarray} 
where $H$ is the entropy function $H(x)=-x \log_2 x-(1-x)\log_2(1-x)$. 
The corresponding asymptotic form for the quantum Gilbert-Varshamov bound
(\ref{gvbound}) is
\begin{equation}
\frac{l}{n}\geq 1-\frac{2t}{n}\log_2 3 -H(\frac{2t}{n}).\label{gvasym}
\end{equation}
As we can see from eq. (\ref{hamasym}), in quantum error correction there is 
an upper bound on the error rate $t/n$ which a code can tolerate. 
In fact, differently from the classical case, where any arbitrary error rate 
can be corrected by a suitable code, in the quantum world the ratio $t/n$ 
cannot be larger than 0.18929 for nondegenerate codes.

\section{System-environment dynamics}
\label{s:dyn}

In order to provide a tangible illustration of some abstract ideas
discussed in the text we have picked up the most popular
quantum-optical model of dissipation commonly used to describe
spontaneous emission. A two level atom, with two energy eigenstates
$\ket{0}$ and $\ket{1}$ separated by $\hbar\omega_0$, interacting with
an environment modelled as a set of quantised harmonic oscillators,
e.g. a set of quantised modes of radiation with frequencies
$\omega_m$. The Hamiltonian of the combined system $H= H_0 +V$
includes both the free evolution of the qubit and the environment. The
free evolution Hamiltonian is given by
\begin{equation}
H_0 = \hbar\omega_0 \proj{1} +\sum_m \hbar\omega_m a^\dagger_m a_m\;,
\end{equation}
where $a_m$ and $a^\dagger_m$ represent the annihilation and
creation operators of the radiation mode of frequency $\omega_m$.
The interaction (in the rotating wave approximation) is described by
\begin{equation}
V = \sum_m \lambda_m \ket{0}\bra{1} a^\dagger_m + \lambda^\star_m
\ket{1}\bra{0} a_m\;,
\end{equation}
where $\lambda_m$ is 
the coupling constant between the qubit and the mode of frequency $\omega_m$.

In order to find the time evolution of the relative states of the
environment $\ket{R_i(t)}$ we need some knowledge about the
qubit-environment interaction. Let us then have a closer look at a
dissipative dynamics in our model of a qubit coupled to a continuum
of field modes or harmonic oscillators.  If all the oscillators in the
environment are in their ground states and the qubit is initially
prepared in state $\ket{\Psi}=\alpha\ket{0}+\beta\ket{1}$ then the
dynamics described by the Hamiltonian $H=H_0+V$ does not affect state
$\ket{0}$. It is state $\ket{1}$ which undergoes a decay. Let us then
consider a case when the initial state of the combined system
(qubit+environment) is
\begin{equation}
\ket{\phi_i}=\ket{1}(\ket{0}_1\ket{0}_2\ldots\ket{0}_f\ldots\ket{0}_{max}),
\end{equation}
meaning the qubit is in state $\ket{1}$ and all the harmonic
oscillators in their ground states $\ket{0}$ (we will denote the state where
all harmonic oscillators are in the ground state as the vacuum 
$\ket{\bf 0}$). Possible final states of the combined system are
\begin{equation}
\ket{\phi_f}=\ket{0}(\ket{0}_1\ket{0}_2\ldots\ket{1}_f\ldots\ket{0}_{max}),
\end{equation}
where the qubit decayed to state $\ket{0}$ and one of the harmonic
oscillators got excited. Let us note that
\begin{equation}
H_0 \ket{\phi_i} = \hbar\omega_0 \ket{\phi_i} ,\quad H_0 \ket{\phi_f}
= \hbar\omega_f \ket{\phi_f},\quad \bra{\phi_f} H_0 \ket{\phi_i} = 0 ,
\quad \bra{\phi_f} V \ket{\phi_i} = \lambda_f .
\label{Ham}
\end{equation}

Let us write $\ket{\phi (t)}$ as
\begin{equation}
\ket{\phi (t)} = c_i(t)e^{-i\omega_0 t}\ket{\phi_i} + \sum_f c_f (t)
e^{-i\omega_f t}\ket{\phi_f}
\end{equation}
which, using our notation from the previous section, implies
$\ket{R_{00}}=\ket{\bf{0}}$, $\ket{R_{01}}=0$, $\ket{R_{10}(t)}=\sum_f
c_f(t)e^{-i\omega_ft}\ket{1_f}$ and $\ket{R_{11}(t)}=c_i(t)e^{-i\omega_0
t}\ket{\bf{0}}$.

In order to find the relevant time dependance we have to solve the
Schr\"odinger equation 
\begin{eqnarray}
i\hbar \dot{c}_i (t) 
& =& \sum_f \lambda_f^\star e^{-i(\omega_f - \omega_0) t} c_f (t)
\label{Schr1}\\
i\hbar \dot{c}_f (t) 
& =& \lambda_f e^{i(\omega_f - \omega_0) t} c_i (t).
\label{Schr2}
\end{eqnarray}
The second equation can be solved formally for $c_f (t)$
\begin{equation}
c_f(t) = - \frac{i}{\hbar} \int_0^t dt' \lambda_f e^{i(\omega_f -
\omega_0) t'} c_i (t')
\label{eqcf}
\end{equation}
and after substituting this expression for $c_f(t)$ in
Eq.(\ref{Schr1}) we obtain

\begin{equation}
\dot {c}_i(t) = - \int_0^t dt' K(t-t') c_i(t'), \qquad K(\tau) =
\frac{1}{\hbar^2}\sum_f|\lambda_f|^2 e^{-i(\omega_f - \omega_0) \tau}.
\end{equation}

It is the function $\lambda_f = \lambda (\omega_f)$ which determines
the character of the evolution.

\begin{itemize} 
\item Parabolic Decay.

At short times, the exponential in $e^{-i(\omega_f - \omega_0)
(t-t')}$ in $K(t-t')$ can be replaced by $1$. This is justified when
$t\ll \frac{1}{\Delta}$, where $\Delta$ is a typical width of the
$\lambda (\omega_f)$ curve. Usually, for a bell-shaped $\lambda
(\omega_f)$ curve the order of $\Delta$ is pretty well approximated by
$\omega_0$. For example if we analyse spontaneous emission in the
optical domain then $\omega_0 = \Delta = 10^{15}\mbox{Hz}$ thus the
short time means here much less than $10^{-15}$ s. The integration in
Eq. (\ref{eqcf}) together with the initial condition $c_i(t=0) = 1$
gives
\begin{equation}
|c_i(t)|^2 = |\bra{\phi_i}\phi(t)\rangle|^2 = 
1 - 2\frac{t^2}{\hbar^2} \sum_f \lambda^2_f.
\end{equation}
The same result can be otained obtained directly by writing
\begin{equation}
\ket{\phi(t)} = e^{-iHt/\hbar}\ket{\phi_i} = (1 - \frac{i}{\hbar} Ht -
\frac{1}{\hbar^2} H^2 t^2 + \ldots)\ket{\phi_i}
\end{equation}
which, together with Eq.(\ref{Ham}) gives
\begin{equation}
|\bra{\phi_i}\phi(t)\rangle|^2 = 1- 2\frac{t^2}{\hbar^2} (\langle H^2
\rangle - \langle H \rangle^2) \ldots = 1-2\frac{t^2}{\hbar^2}\sum_f
\lambda^2_f + \ldots
\end{equation}
Thus for short times the decay is always parabolic. Let us mention in
passing that from a purely mathematical point of view we have assumed
here that expression $(\langle H^2\rangle -\langle H\rangle^2) =
\sum_f \lambda^2_f $, i.e. the variance of the energy in the initial
state $\ket{\phi_i}$, is finite. Needless to say in reality it is
always finite but there are mathematical models in which, due to
various approximations, this may not be the case (e.g. the Lorentzian
distribution which has no finite moments).

\item Exponential Decay.

Expression $|\lambda_f|^2 e^{-i(\omega - \omega_0) \tau}$ viewed as a
function of $\omega_f -\omega_0$ oscillates with frequency $1/\tau$
whereas $\lambda_f = \lambda (\omega_f)$ varies smoothly in the
frequency domain. Again taking $\Delta$ as the typical width of the
$\lambda (\omega_f)$ curve for $\tau >> 1/ \Delta$ the sum in
$K(\tau)$ averages out to zero. This allows to substitute $c_i(t)$ for
$c_i(t')$ in Eq.(\ref{Schr1}) which gives
\begin{equation}
\dot {c}_i(t) \approx -c_i(t)\int_0^t d\tau K(\tau) \approx
-c_i(t)\int_0^\infty d\tau K(\tau).
\end{equation}
Now we can calculate $\int_0^\infty d\tau K(\tau)$ using the identity
\begin{equation}
\int_0^\infty d \tau e^{i\omega\tau} = \lim_{\epsilon \rightarrow 0^+}
\int_0^\infty d \tau e^{i(\omega + i\epsilon)\tau} = \lim_{\epsilon
\rightarrow 0^+}\frac{i}{\omega+i\epsilon} = i\mbox{\cal
P}\frac{1}{\omega} + \pi \delta (\omega).
\end{equation}
It gives
\begin{equation}
\int_0^\infty d\tau K(\tau) = \frac{\gamma}{2} + i \delta, 
\quad \frac{\gamma}{2} =
\frac{\pi}{\hbar^2} |\lambda (\omega_f=\omega_0)|^2, \quad \delta =\mbox{\cal
P}\sum_f\frac{|\lambda_f|^2}{\omega_0-\omega_f}.
\end{equation}
Incorporating the energy shift $\hbar \delta$ into the modified energy
separation $\hbar (\omega_0+\delta)$ we finally obtain
\begin{equation}
\dot{c}_i(t) = -\frac{\gamma}{2} c_i(t)\qquad \mbox{that is} \qquad
c_i(t) = e^{-\frac{\gamma t}{2}}
\end{equation}
and consequently
\begin{equation}
c_f(t)=\frac{\lambda_f}{\hbar} \frac{1-e^{i(\omega_f-\omega'_0 +
i\gamma/2)t}}{\omega_f-\omega'_0 + i\gamma/2}
\end{equation}

\end{itemize}

Let us now go back to the language introduced in section \ref{s:int}. 
The states of the environment $\ket{R_0(t)},\\ \ket{R_1(t)}, \ket{R_2(t)}$
and $\ket{R_3(t)}$ in the present context take the explicit form
\begin{eqnarray}
\ket{R_0(t)}&=&\frac{1}{2}[1+c_i(t)e^{-i\omega_0 t}]\ket{\bf 0}\;,
\label{R0}\\
\ket{R_1(t)}&=&\frac{1}{2}\sum_f c_f(t) e^{-i\omega_f t}\ket{1}_f\;,
\label{R1}\\
\ket{R_2(t)}&=&-\frac{1}{2}\sum_f c_f(t) e^{-i\omega_f t}\ket{1}_f\;,
\label{R2}\\
\ket{R_3(t)}&=&\frac{1}{2}[1-c_i(t)e^{-i\omega_0 t}]\ket{\bf 0}\;.
\label{R3}
\end{eqnarray}
By formula (\ref{fidelity}), the fidelity of this process is given by
\begin{eqnarray}
F(t)&=&\bra{R_0(t)} R_0(t)\rangle  + \bra{R_3(t)} R_3(t) \rangle
-2\mbox{Re}\bra{R_0(t)}R_3(t)\rangle\nonumber\\
&=&|c_i(t)|^2\;.                                                        
\end{eqnarray}
Therefore, the fidelity in the case of a parabolic decay takes the form
\begin{eqnarray}
F_{par}(t)= 
1 - 2\frac{t^2}{\hbar^2} \sum_f \lambda^2_f\;,
\end{eqnarray}
while in the case of an exponential decay it has the exponential form
\begin{eqnarray}
F_{exp}(t)= e^{-\gamma t}\;.
\end{eqnarray}

\section{Benefits of quantum error correction}
\label{s:beyond}

In order to get an idea about the efficiency of quantum error correction, we will now discuss
an explicit construction of the single error-correcting five qubit code. 
The initial state of the qubit $\alpha\ket{0}+\beta\ket{1}$ is encoded in state
$\alpha\ket{C_0}+\beta\ket{C_1}$, where \cite{lafl}
\begin{eqnarray}
\ket{C_0}&=&\ket{00010}+\ket{00101}-\ket{01011}+\ket{01100}
+\ket{10001}-\ket{10110}-\ket{11000}-\ket{11111}\\
\ket{C_1}&=&\ket{00000}-\ket{00111}+\ket{01001}+\ket{01110}
+\ket{10011}-\ket{10100}+\ket{11010}-\ket{11101}.
\end{eqnarray}
(To see the benefits of quantum error correction we do not need to use
the explicit form of the code, we wrote it down here for those
curious readers who may want to play with quantum error correcting codes.)  
These encoded
states are chosen in such a way that conditions (\ref{orth}) are
satisfied. Since this code can correct any type of error affecting one
qubit, it is suitable for protecting quantum states against 
spontaneous emission. We notice
that the spontaneous emission process described in Sect.  \ref{s:dyn},
unlike decoherence, involves both phase and amplitude errors and
therefore it cannot be successfully defeated with less than five
bit codes.

The probability that the state undergoes exponential decay in the presence 
of spontaneous emission is approximately given by 
\begin{eqnarray}
P_{dec}(t)=1-F_{exp}(t)=1-e^{-\gamma t}\;.
\end{eqnarray}
If we assume that the five qubits decay independently from each other, the
probability that none of them decays is given by
\begin{eqnarray}
P_{no\;dec}(t)=e^{-5\gamma t}\;
\end{eqnarray}
while the probability that only one of them decays is
\begin{eqnarray}
P_{1\;dec}(t)=e^{-4\gamma t}(1-e^{-\gamma t})\;.
\end{eqnarray}
Since by construction the above error correction scheme corrects perfectly
the encoded state when only one of the qubits is affected, the fidelity
of reconstruction of the state after the error correction is at least as
high as the probability of having at most one qubit decay during the process,
that is
\begin{eqnarray}
F_{ec}(t)\ge P_{no\;dec}(t)+ 5P_{1\;dec}(t)=e^{-4\gamma t}(5-4e^{-\gamma t}).
\end{eqnarray}
In order to have a successful error correction the such fidelity must be greater than the fidelity
$F_{exp}(t)$ corresponding to a single qubit in the absence of error correction. This is true when
the decay probability
$P_{dec}(t)$ is much smaller than one, namely when the correction procedure   is applied at times
$t\ll 1/\gamma$. Actually, for $t\ll 1/\gamma$ the fidelity of reconstruction after error correction
is bounded by
\begin{eqnarray}
F_{ec}(t)\ge 1-10\gamma^2 t^2 + O(t^3)\;,
\end{eqnarray}
namely it has parabolic form, while the single qubit decay probability is
\begin{eqnarray}
P_{dec}(t)\simeq 1-\gamma t\;.
\end{eqnarray}

\section{Concluding remarks}

Research in quantum error correction in its all possible variations has become
vigorously active and any comprehensive review of the field must be obsolete as soon
as it is written. Here we have decided to provide only some very basic knowledge,
hoping that this will serve as a good starting point to enter the field. The reader
should be warned that we have barely scratched the surface of the current activities
in quantum error correction neglecting topics such as group theoretical ways of
constructing good quantum codes \cite{GF4}, concatenated codes 
\cite{knill}, quantum fault tolerant computation \cite{divi-shor} and many  others.
Many interesting papers in these and many related areas can be found at the Los
Alamos National Laboratory e-print archive  (http://xxx.lanl.gov/archive/quant-ph).

  This work was supported in part by the European TMR Research Network
  ERP-4061PL95-1412, the TMR Marie Curie Fellowship Programme, 
  Hewlett-Packard, The Royal Society London and Elsag-Bailey, 
  a Finmeccanica Company.

\end{document}